\documentclass [12pt,a4paper]{article}    %
\usepackage{times}

\DeclareFontFamily{OT1}{times}{}
\DeclareFontShape {OT1}{times}{m }{n }{ <-> ptmr }{}
\DeclareFontShape {OT1}{times}{bx}{n }{ <-> ptmb }{}
\DeclareFontShape {OT1}{times}{m }{it}{ <-> ptmri}{}
\DeclareFontShape {OT1}{times}{bx}{it}{ <-> ptmbi}{}
\usepackage{amsmath}
\usepackage{amsfonts}
\usepackage{amssymb}
\usepackage{latexsym}

\numberwithin{equation}{section}

\newcommand{\cl}{C \kern -0.1em \ell_\pi} 
\newcommand{\DEF}{:=}                 

\newcommand{\CON}{\overline}          

\newcommand{\scal}{\circ}             
\newcommand{\vect}{\wedge}            

\newcommand{\SCA}{\langle}            
\newcommand{\LAR}{\rangle}            
\newcommand{\VEC}{\vec{\kern +.1em[}} 
\newcommand{\TOR}{\vec{\kern +.2em]}} 
\newcommand{\BRA}{\langle\kern -.2em\langle} 
\newcommand{\KET}{\rangle\kern -.2em\rangle} 

\newcommand{\e  }{\operatorname{e}}   
\newcommand{\Oh}{\tfrac{1}{2}}        

\newcommand{\Q}{[\hspace{1.mm}]} 
\newcommand{\A}{(\hspace{.5mm})} 

\begin{document}

\title{\bf\vspace{-2.5cm} On the physical interpretation of\\ singularities 
                                in Lanczos-Newman\\ electrodynamics }

\author{
         {\bf Andre Gsponer}\\
         {\it Independent Scientific Research Institute}\\ 
         {\it Oxford, OX4 4YS, England}
       }

\date{ISRI-04-04.36 ~~ November 9,  2008}

\maketitle

\begin{abstract}

We discuss the physical nature of elementary singularities arising in the complexified Maxwell field extended into complex spacetime, i.e., in Lanczos-Newman electrodynamics, which may provide a possible link between elementary particle physics and general relativity theory.  We show that the translation of the world-line of a bare (e.g., spinless) electric-monopole singularity into imaginary space is adding a magnetic-dimonopole component to it, so that it can be interpreted as a pseudoscalar pion-proton interaction current, consistent which both charge-independent meson theory and zero-order quantum chromodynamics.  On the other hand, the interaction current of an electric-monopole intrinsic-magnetic-dipole singularity characteristic of a Dirac electron is obtained by another operation on the world-line, which however does not seem to have a simple geometric interpretation.  Nevertheless, both operations can be given a covariant interpretation, which shows that the corresponding interactions necessarily arise on an equal footing, and therefore provides a connection between elementary particles and singularities in general relativity.

\end{abstract}	

%
%
%
%
%
%
%
%
%
%
%

\section{Introduction} 
\label{int:0}

In 2004 Ezra Newman showed that a generalized form of the Li\'enard-Wiechert fields, solution to Maxwell's equations extended into complex spacetime, may correspond to the world-lines of a charged particle having a finite magnetic-dipole moment and a gyromagnetic ratio of two \cite{NEWMA2004-}.  Since such a particle would be a Dirac electron, it is of great interest that earlier work done in this direction could be confirmed and expanded \cite{NEWMA1973-, NEWMA2002-}.

However, as stressed by Newman, there are great difficulties, both \emph{technical} because of the complicated mathematics inherited from the general relativity considerations which gave rise to his new approach to the problem of electricity in curved spacetime, and \emph{physical} because the interpretation of the singularities associated with arbitrary world-lines in complex spacetime is not trivial, even in flat Minkowski space.\footnote{According to Newman the singularities of fields may have no fundamental significance, and (other than as approximation techniques) they may not have any role to play in general relativity [Ezra Newman, private communication, 29 October 2008].}

The focus of this paper is on the second class of difficulties, mainly because they are closest to the author's domain of expertise.  This paper is therefore of a cross-disciplinary kind, implying that since its goal is to interpret Newman's world-lines in the perspective of their possible relevance to elementary particle physics, we use a language and a methodology different from his.  In particular, instead of congruences of `world-lines' in complexified Minkowski space, we deal with `poles' in the complexified Maxwell field,\footnote{Such poles might be simple spherically-symmetric distributions such as the Coulomb-like potential of electric or magnetic monopoles; or more complicated distributions such as, for example, the axially-symmetric magnetic-dipole of electrons; as well as combinations of such poles, e.g, the electric-monopole magnetic-dipole combination characteristic of a Dirac electron, Eq.\,\eqref{int:3}.}  
i.e., with `point-like' singularities which can be put in one-to-one correspondence with world-lines by means of special or general Lorentz transformations.   Therefore, in most of the paper, we work in the non-relativistic limit, that is in the rest-frame of the singularities, what is appropriate to study their distinctive features and to interpret them.

   In other words, we assume that the elementary particles known from experiment could correspond to poles in complexified spacetime, and that at least the most common of them --- namely the electrons, nucleons, and pions --- could in some appropriate low-energy limit\footnote{By low-energy we mean energies $\mathcal{E}$ that are small in comparison to the mass of the heaviest particles under consideration, i.e., the nucleons.  Therefore $\mathcal{E} \ll 939 \text{MeV} \approx 1 \text{GeV}$, i.e., energies that are very low in comparison to current elementary particle research, which is done using accelerators with energies on the order of 1000 GeV.  For this reason, the `parton aspect' of elementary particles (i.e., their modeling in terms of quarks, gluons, etc.), which is the main focus of present-day high-energy physics, is not essential in the energy domain considered in this paper.  Nevertheless, in Sec.\,\ref{qua:0}, we will show how the main results of this paper can be expressed in terms of these concepts.}  
correspond to poles in the Maxwell and Dirac fields expressed over complexified spacetime.\footnote{The assumption that in the low-energy limit the nucleons and pions are `whole particles,' just like electrons and photons, is in good agreement with experiment, and gives a satisfactory description of essentially all of atomic and nuclear physics. In particular, the contemporary theory of nuclear forces is still essentially that of the 1950s, with the difference that the constraints coming from its high-energy limit (in which quark and gluon effects dominate) is now better understood \cite{SEROT1997-, MACHL2003-}.  However, essentially the same constraints (from isospin and chiral symmetry) arise from Lanczos's equation of 1929 and its 1932 generalization by Einstein and Mayer \cite{GSPON2001-}.}

   Fortunately, this idea is not new, and was systematically explored in 1919 already by Cornelius Lanczos in his PhD dissertation \cite{LANCZ1919-, HURNI2004-, GSPON1998A}.  In this dissertation he postulated that electrons and protons could be singularities of the Maxwell field expressed in the formalism of complex quaternions, and showed that this formalism is most appropriate to study such singularities when they are analytically continued into complexified Minkowski space.  In particular, to deal with the problem of infinities which necessarily arises when integrating quantities associated with even the most simple singularities, Lanczos had the idea of shifting the origin of the Coulomb potential by a small imaginary amount into imaginary 3-space, i.e., 
\begin{equation}\label{int:1} 
   \frac{e}{|\vec{r}\,|} \rightarrow  \frac{e}{|\vec{r} +i\vec{\lambda}\,|},
\end{equation}
so that the Coulomb singularity is no more a point but a circle.  This `circle electron' --- which as observed by Lanczos has the nice property of yielding a finite Hamiltonian function --- will be rediscovered by others, for instance by Newman who saw in it a possibility to endow a bare Coulomb monopole with a magnetic moment \cite{NEWMA1973-, NEWMA2002-}.

   However, since the Coulomb potential $\varphi_{\text{monopole}} = e/r$ where $r=|\vec{r}\,|$ is just a \emph{scalar} in the rest-frame of the electron, the only effect of shifting its origin by an imaginary amount is to add an imaginary part to the potential, and therefore to make it complex instead of real.  Indeed
\begin{equation}\label{int:2} 
 \frac{e}{|\vec{r}\,|} \rightarrow  \frac{e}{|\vec{r} +i\vec{\lambda}\,|}
    = \frac{e}{|\vec{r}\,|}
   + \Bigl( \frac{e}{|\vec{r} +i\vec{\lambda}\,|} - \frac{e}{|\vec{r}\,|} \Bigr)
    \approx \frac{e}{r} - ei\frac{\vec{\lambda}\cdot\vec{r}\,}{r^3}
    \in \mathbb{C}.
\end{equation}
While this complexified scalar potential leads to an electromagnetic field having some of the characteristics of a magnetic dipolar field, it is nevertheless unsuitable to yield the exact field of an \emph{intrinsic magnetic dipole:}  Such a field derives from the \emph{vector} potential $\vec{A}_{\text{dipole}} = (\vec{\mu}\times \vec{r})/r^3$, not from a scalar potential.  Thus, the interpretation of the singularities obtained by displacing a world-line into complexified spacetime is more complicated than anticipated by Newman.  

   As will be shown in Section $\ref{phy:0}$,  what is actually generated by means of a Lanczos-Newman's infinitesimal `imaginary scalar translation' is a pseudoscalar field which can be associated with the pion field of a nucleon.  The intrinsic magnetic dipole field characteristic of a Dirac electron is obtained by a different transformation:  An infinitesimal\footnote{The magnitudes of the transformations considered in this paper are infinitesimal in the sense that they correspond to lengths characteristic of elementary particles, i.e., to displacements characterized by their Compton wave-length, $\hbar c/mc^2$, which are on the order of~$10^{-15}$ m.} `imaginary quaternion operation' which is adding a vector part to Coulomb's scalar potential so that it becomes the 4-potential of the electromagnetic field of a Dirac electron which, in its rest-frame, is the biquaternion 
\begin{equation}\label{int:3}
     A_{\text{e}} = \varphi_{\text{monopole}} - i \vec{A}_{\text{dipole}} 
         = \frac{e}{r} - i\frac{\vec{\mu} \times \vec{r} }{r^3} 
           \in \mathbb{B}.
\end{equation}

  In order for such operations to be possible and meaningful it is necessary to write Maxwell's field as a complex field over complex spacetime, and to correctly associate the poles and sources of that field to elementary particles.  This requires a theory in which both Maxwell's and Dirac's equations are expressed over the same space, i.e., everyday's Minkowski spacetime extended to complexified spacetime by analytic continuation, and a common formalism which can handle the photon, electron, and possibly other particle's fields consistently.  Such a formalism has been developed by Cornelius Lanczos in 1919 for Maxwell's equations \cite{LANCZ1919-, HURNI2004-, GSPON1998A, GSPON2004E}, and in 1929 for Dirac's  equations \cite{LANCZ1929A, LANCZ1929B, LANCZ1929C, GSPON1998B}.

    Lanczos's formalism is briefly reviewed in Sections \ref{max:0} and \ref{dir:0}.\footnote{Sections \ref{max:0} and \ref{dir:0} can be skipped at first reading: Their purpose is to show that Maxwell's and Dirac's theory can be expressed in complex spacetime, and to derive the magnitude of the intrinsic magnetic moment of a Dirac particle, Eq.\,\eqref{dir:7}.} It is based on Hamilton's biquaternion (i.e., complexified quaternion) algebra $\mathbb{B}$, which is defined by the non-commutative product\footnote{The quaternion product is a direct generalization of the complex number product, which was defined by Hamilton as $(a+iA)(b+iB)=ab-AB+iaB+ibA$.}
\begin{equation}\label{int:4}
(a+\vec{A})(b+\vec{B})=ab-\vec{A}\cdot\vec{B} 
      + a\vec{B} + b\vec{A} + \vec{A}\times\vec{B} .
\end{equation}
where $a, b \in \mathbb{C}$ are complex scalars and $\vec{A}, \vec{B} \in \mathbb{C}^3$ are complex vectors.  Using this unique multiplication rule, which provides a very straightforward and general formulation of special relativity for both integer and half-integer spin (i.e., bosonic and fermionic) fields, it is possible to formulate not just Maxwell's and Dirac's theories, but the whole of classical and quantum theory in a concise and consistent way which avoids many of the complications inherent in the standard formalism \cite{GSPON1993-,GSPON1994-}.  Moreover, while it is well known that basic 4-vectors such a the 4-position $\mathcal{X} = ict + \vec{x}$, or the 4-momentum $\mathcal{P} = \mathcal{E} - ic\vec{p}$, can most naturally be written as biquaternions, it is important for the purpose of this paper that the fields and singularities associated with elementary particles can also be expressed as biquaternions, e.g., \eqref{int:3}.

   Section \ref{sin:0} is a simplified discussion of the elementary point-like singularities which appear in a biquaternion field theory, namely the monopole, dipole, and dimonopole singularities.  Its main objective is to show that a magnetic dimonopole (a scalar-potential singularity) produces the same magnetic field as an intrinsic magnetic dipole (a vector-potential singularity), \emph{except at the origin where the two singularities produce different delta-function-like fields}. Therefore, despite their similarity at $r\neq 0$, the dipole and dimonopole fields correspond to entirely different physical particles.

   In Section \ref{phy:0} we first show that the Lanczos-Newman translation leading to the `circle electron' singularity yields a pseudoscalar-potential rather than the vector-potential of Dirac's electron magnetic-dipolar field, and then we derive the operation which yields the correct potential for that field.

  In Section \ref{qua:0} we show that the interpretation of the Lanczos-Newman dimonopole singularity as a strong-interaction potential is consistent with the quark model and quantum chromodynamics.

    Finally, in Section \ref{cov:0} we give a covariant interpretation of the scalar and quaternion operations which lead from the bare electric-monopole potential of classical electrodynamics to the proton-pion and electron-photon interactions potentials of quantum theory.

\section{Lanczos's formulation of Maxwell's equations} 
\label{max:0}

The fundamental quantity in Lanczos's biquaternion\footnote{Hamilton introduced the prefix `bi-' to indicate that some quantity is complexified.  For example, Hamilton called a complex number a `biscalar.' We denote quaternion conjugation as $\CON{Q}$ and complex conjugation as $Q^*$, and we define biconjugation as $Q^+\DEF\CON{Q}\,^*$.}  formulation of Maxwell's equations is the Silberstein-Conway electromagnetic field \emph{bivector} $B$ which combines the electric and magnetic fields in a single complex vector (which transforms as a 6-vector, i.e., $B \rightarrow \mathcal{L}^* B \mathcal{L}^+$ in a Lorentz transformation\footnote{A general Lorentz transformation, i.e., a rotation followed by a boost, or vice versa, is simply a unit biquaternion $\mathcal{L}\in \mathbb{B}$ such that $\mathcal{L}\CON{\mathcal{L}}=1$.})
\begin{equation}\label{max:1}
          B \DEF \vec{E} + i \vec{H} .
\end{equation}
This bivector should not be confused with the Faraday-Kilmister electromagnetic field \emph{tensor}, which in biquaternions corresponds to the linear function\footnote{The symbol $\A$ is embracing the argument of a function, whereas the symbol $\Q$ is a place-holder indicating where this argument goes in the function, taking non-commutativity of quaternions into account.}
\begin{equation}\label{max:2}
        F\A = \tfrac{1}{2} ( \Q B - B^+    \Q ) .
\end{equation}

Maxwell's equations are then  \cite{LANCZ1919-, GSPON1993-}
\begin{equation}\label{max:3}
       \CON{\nabla} \vect A = B,  ~~~   ~~~  \nabla B = -C,
\end{equation}
where $A$ is the 4-potential (which transforms as a 4-vector, i.e., $A \rightarrow \mathcal{L} A \mathcal{L}^+$ in a Lorentz transformation)
\begin{equation}\label{max:4}
       A \DEF \varphi - i\vec{A} ,
\end{equation}
and $C$ the non-rationalized source 4-current density, which is related to the usual rationalized charge-current density
\begin{equation}\label{max:5}
       J \DEF \rho -i\vec{j} ,
\end{equation}
by the equation
\begin{equation}\label{max:6}
                 C \DEF 4 \pi J  .
\end{equation}

In equation \eqref{max:3} the operator $\nabla$ is the four-dimensional generalization of Hamilton's differential operator $\vec\nabla$, i.e.,
\begin{equation}\label{max:7}
            \nabla\A  = \frac{\partial}{\partial ict}\Q + \vec\nabla\Q,
\end{equation}
and the operator $\vect$ means that the scalar part is discarded so that $B$ is a bivector as it should be according to \eqref{max:1}.  By keeping all parameters real, and interpreting $i$ as a device to get Minkowski's metric with Hamilton's real quaternion product, equation \eqref{max:3} can be seen as just another way of rewriting Maxwell's equations in a form which is equivalent to the more standard tensor, vector-algebra, or differential-form formulations.  However, if we follow the suggestion of Lanczos and Newman, and thus accept the complexification of the potential, fields, and sources, as well as any kind of transformation on or between their scalar and vector parts, including (possibly complex) transformations of the coordinates, we get generalized types of electric and magnetic fields which are considered as `unphysical' in the standard interpretation of Maxwell's field as a real field over real Minkowski space.\footnote{In the biquaternion formalism the standard Maxwell field corresponds to the restriction to bireal potentials, i.e., such that $A=A^+$, which have 4 real components instead of 8.}

In this paper we interpret these generalized biquaternionic fields in terms of experimentally known elementary particle fields.  The reason why such an interpretation makes sense is that, as discovered by Lanczos \cite{LANCZ1929A, LANCZ1929B, LANCZ1929C}, there is a very close relationship between the complexified Maxwell field and the standard Maxwell, Dirac, and Proca fields, which can all be interpreted as distinct superpositions of a complex `Lanczos field' \cite{GSPON2001-, GSPON1998B, GSPON2002-}.

\section{Lanczos's formulation of Dirac's equation} 
\label{dir:0}

Soon after the publication of Dirac's famous paper in which the quantum field of the electron was expressed in an abstract space using four-dimensional matrix algebra, Lanczos showed that Dirac's equation could be expressed in everyday's spacetime just like Maxwell's equations, and that working solely with vector algebra in its original form, namely biquaternions, one could dispense of any sophisticated formalism such as the cumbersome `Dirac matrices.' In fact, as the titles of his three papers suggest, \emph{The tensor analytical relationships of Dirac's equation} \cite{LANCZ1929A}, \emph{The covariant formulation of Dirac's equation} \cite{LANCZ1929B}, and \emph{The conservation laws in the field theoretical representation of Dirac's theory} \cite{LANCZ1929C}, Lanczos was motivated by showing that Dirac's equation had a perfectly sensible and straightforward tensor interpretation, just like Maxwell's, Einstein's, and all other fundamental equations of physics. 

   Lanczos showed that the Maxwell and Dirac fields are just two distinct superpositions of a more general field, which he could not interpret.  But with todays hindsight that interpretation is not difficult, and it readily seen that the full Lanczos field is perfectly suited to express the physics of the interactions of photons and electrons and their weak-isospin partners, namely the weak-interaction intermediate-bosons and the neutrinos, exactly as given by today's `Standard Model' of electro-weak interactions \cite{GSPON2001-}.  Moreover, as shown by Feza G\"ursey, Lanczos's field can easily accommodate strong interactions, at least in the low-energy limit in which these interactions are given by the charge-independent theory based on pseudoscalar pion fields \cite{GURSE1958-}.

   In Lanczos's representation the Dirac field is a biquaternion $D \in \mathbb{B}$ satisfying the \emph{Dirac-Lanczos equation} \cite[eq.\,63]{LANCZ1929A}
\begin{equation}\label{dir:1}
   \CON{\nabla} D =  \frac{1}{\ell} D^* i\vec{\nu} +
                \frac{\alpha}{e} \CON{A} D   i\vec{\nu}  ,
\end{equation}
and transforming as a 4-spinor, i.e., $D \rightarrow \mathcal{L} D$ in a Lorentz transformation. Here $A$ is the electromagnetic 4-potential \eqref{max:4}, $\alpha = e^2/\hbar c$ the electromagnetic coupling constant,  $\ell = \hbar/ mc$ the Compton wave-length associated with the mass $m$, and $\vec{\nu}$ an arbitrary unit vector.\footnote{This unit vector defines an axis relative to which quantities such as the spin are quantized.  Its arbitrariness corresponds to the freedom in the choice of a set of anticommuting matrices in Dirac's representation. The presence of this non-scalar factor in the Dirac-Lanczos equation \emph{insures} that the field $D$ is of spin-$\Oh$ exclusively, and therefore cannot be of spin-$1$ as Maxwell's for example.}   While Lanczos's form is fully equivalent to other forms of Dirac's equation, its major virtue  is the presence of a complex conjugation on the right hand side, which explicitly shows that Dirac's and Maxwell's equations are fundamentally different because that conjugation operation \emph{implies} that Dirac's field is fermionic, and therefore obeying Pauli's exclusion principle, whereas Maxwell's field is bosonic \cite{GSPON2002-}.

For the purpose of this paper we do not need more than Dirac's bilinear covariant quantities,\footnote{In Lanczos's formalism the 16 bilinear covariant quantities of Dirac's theory (two 4-vectors, one invariant biscalar, and one 6-vector) can immediately be written down in explicit form without the need of operators as in Dirac's matrix formalism.} i.e.,  Dirac's \emph{conserved 4-current}\footnote{In this paper we write Dirac's conserved probability density 4-vector as an electromagnetic current density, which is why we include the electric charge $e$ in its definition. Probability-current and electromagnetic charge-current density conservation are then expressed by the identity $\CON\nabla \cdot C = 0$. The use of the same symbol $C$ for Dirac's current $\eqref{dir:2}$ and for the source current in Maxwell's equations $\eqref{max:3}$ is consistent because this is precisely how the Dirac and Maxwell equations are coupled in field theory, namely by having common $A$ and $C$.} $C$ and \emph{spin pseudo-4-vector} $\Sigma$ 
\begin{equation}\label{dir:2}
                     C = eDD^+ ,   ~~ ~~  ~~ \Sigma =  D \vec\nu D^+,
\end{equation}
and Dirac's \emph{invariant scalar} $s$ and \emph{spin 6-vector} $S$
\begin{equation}\label{dir:3}
                        s = D\CON{D} ,    ~~ ~~  ~~ S = D \vec\nu \CON{D};
\end{equation}
as well as \emph{Gordon's decomposition} of $C$, which is an algebraic identity deriving from \eqref{dir:1} and \eqref{dir:2} that we write in the general form given by G\"ursey \cite[p.\,50--52]{GURSE1956B}, i.e.,\footnote{Gordon's decomposition in its traditional form is restricting the terms on the right-hand-side to their bireal (i.e., `Hermitian') part, which has the effect of specializing the Dirac equation to electromagnetic interactions, even though Dirac particles such as the proton also have non-electromagnetic interactions.}
\begin{equation}\label{dir:4}
          \frac{1}{\ell}C = s\alpha  A
              -\varepsilon(\nabla)
              -\frac{1}{2}\mathcal{M}(\nabla) .
\end{equation}
In this decomposition the first two terms correspond to the \emph{convection current}, which would exist even if the electron had no spin, and the last one to the \emph{polarization current}, which is due to spin. The linear function\footnote{The underline specifies the range of action of the operator inserted in the square brackets, and the angle brackets mean taking the scalar part.}
\begin{equation}\label{dir:5}
     \varepsilon\A =   e \underline{ \Q \SCA D } i\vec\nu  \CON{D}  \LAR ,
\end{equation}
is the \emph{electric convection scalar}, and the linear function
\begin{equation}\label{dir:6}
     \mathcal{M}\A = e\underline{D i\vec\nu\CON{D}\Q} ,
\end{equation}
the \emph{electromagnetic moment bivector}.  Comparing \eqref{dir:4} and \eqref{dir:6}, we see the magnetic moment of a Dirac particle is given by
\begin{equation}\label{dir:7}
    \vec{\mu} = \frac{1}{2} \ell e \vec{\nu} = \frac{e \hbar}{2 m c} \vec{\nu}.
\end{equation}

\section{Elementary singularities in polar coordinates} 
\label{sin:0}

In this section we consider the simplest point-like singularities which occur in three-dimensional space, and which for convenience we assume to be located at the origin of a polar coordinate system.  This requires a general method to deal with functions in the variable $\vec{r} \in \mathbb{R}^3$ that are singular at the origin of a spherical coordinates system $\{r, \theta, \phi \}$ where $r = |\vec{r}\,|$.  Because the derivative of the absolute value $|\vec{r}\,|$ is \emph{discontinuous} at $\vec{r}=0$, this method must rely on distribution theory to take into account the $\delta$-singularity which arises when differentiating $|\vec{r}\,|$.  An example of such a method is Tangherlini's algorithm for differentiating at the origin in polar coordinates \cite[p.\,511--513]{TANGH1962-}. 

  In this paper, however, we will use a more general method which has the advantage of being suitable for making linear as well as nonlinear calculations with singular functions.  This method is presented in simplified form in \cite{GSPON2004D,GSPON2006B} and in a fully rigorous manner in \cite{GSPON2008B}.  For the purpose of the present paper this method consists of replacing the Coulomb potential $e/|\vec{r}\,|$ by the potential $e\Upsilon(r)/r$ where the nonlinear generalized function $\Upsilon(r)$ has the properties
\begin{equation} \label{ele:1}
       \frac{d}{dr}\Upsilon(r) = \delta(r),
       \qquad \text{and} \qquad
       \int_0^\infty dr~ \Upsilon(r)F(r) =  \int_0^\infty dr~ F(r),
\end{equation}
which are similar to the properties of Heaviside's step function, and where $F(r) \in \mathcal{C}^\infty$ is any smooth function of $r$.

\subsection{Monopole singularity} 
\label{mon:0}

Starting from the Coulomb potential of a point-charge,  
\begin{equation} \label{mon:1}
       \varphi_m(\vec r) \DEF e\frac{1}{r}\Upsilon(r) ,
\end{equation}
which may be either an electric-monopole of charge $e$, or a magnetic-monopole if $e$ is replaced by $ie$, we get the field 
\begin{equation} \label{mon:2}
       \vec E_m(\vec r) = -\vec \nabla \varphi_m
                      =  e\frac{\vec r}{r^3}\Upsilon(r) 
                      -  e\frac{\vec r}{r^2}\delta(r) ,
\end{equation}
and the rationalized source charge distribution
\begin{equation} \label{mon:3}
       4\pi \rho_m(\vec r) =   \vec \nabla \cdot \vec E_m 
                           =  e \frac{1}{r^2} \delta(r) ,
\end{equation}
which upon integration yields the \emph{charge} of the source
\begin{equation}\label{mon:4}
     q =  \iiint d^3\Omega ~ \rho_m(\vec r) = e  .
\end{equation}

\subsection{Dipole singularity} 
\label{dip:0}

The dipole singularity, which through extensive experimental verification is found to very precisely characterize the intrinsic magnetic dipole moment of elementary particles such as the electron, is given by the \emph{vector} potential
\begin{equation}\label{dip:1}
   \vec A_d(\vec r) \DEF
             \frac{\vec{\mu} \times \vec{r} }{r^3}\Upsilon(r),
\end{equation}
where $|\vec{\mu}\,|$ has the dimension of a charge times a length. The calculation of the magnetic field is straightforward.  We get
\begin{equation}\label{dip:2}
    \vec H_{d}(\vec r) =  
        \Bigl( 3\frac{\vec  r}{r^5}(\vec\mu\cdot\vec r)
              - \frac{\vec\mu}{r^3} \Bigr)  \Upsilon(r)
 +        \frac{\vec{r} \times (\vec\mu\times\vec{r})}{r^4}\delta(r) .
\end{equation}
The first term in this expression is well-known, but the one with the $\delta$-function is generally ignored. However, when integrated over 3-space, this second term gives the contribution \cite[p.\,184]{JACKS1975-}
\begin{equation}\label{dip:3}
       \iiint d^3\Omega ~
       \delta(r)~ \frac{\vec{r} \times (\vec\mu\times\vec{r})}{r^4}
     = \frac{8\pi}{3} \vec\mu ,
\end{equation}
which is essential in calculating the hyperfine splitting of atomic states \cite{JACKS1977-}. 

We can now calculate the sources.  As expected, the magnetic charge density is zero
\begin{equation}\label{dip:4}
      4\pi \rho_{d}(\vec r) = \vec\nabla \cdot \vec H_{d}(\vec r) = 0 ,
\end{equation}
while the rationalized current density is
\begin{equation}\label{dip:5}
    4\pi \vec j_{d}(\vec r) = \vec\nabla \times \vec H_{d}(\vec r)
        =  3 \frac{\vec{\mu} \times \vec{r} }{r^4} \delta(r) . 
\end{equation}
Using this current density we can now calculate the \emph{magnetic moment} by means of the standard expression \cite[p.\,181]{JACKS1975-} to get
\begin{equation}\label{dip:6}
            \vec m = 
 \frac{1}{2} \iiint d^3\Omega ~ \vec r \times \vec j_{d}(\vec r)
                   =  \vec \mu .
\end{equation}
Therefore, although there are actually no `circulating currents' in the point-like distribution \eqref{dip:5}, the magnetic moment calculated with the formula derived for a localized current distribution gives the correct answer.

\subsection{Dimonopole singularity} 
\label{dim:0}

The dimonopole singularity corresponds to the field produced by two electric  (or magnetic) monopoles of opposite charge separated by an infinitesimal distance $|\vec \lambda|$. The potential for such a field is therefore
\begin{equation}\label{dim:1}
   \varphi_{dm}(\vec r) \DEF \frac{e}{|\vec{r}\,|}
                           - \frac{e}{|\vec r + \vec\lambda|} .
\end{equation}
At large distance, or at vanishingly small separation $\vec \lambda$, we can take for this potential the first term of the Taylor development, i.e.,
\begin{equation}\label{dim:2}
   \varphi_{dm}(\vec r) \approx
    - e(\vec\lambda \cdot \vec \nabla) \frac{1}{r} \Upsilon(r) =
      e(\vec\lambda \cdot \vec r) \frac{1}{r^3} \Upsilon(r)  .
\end{equation}
From there on it is possible to calculate the field and the source by either recursively applying the gradient operator on \eqref{dim:2}, or by applying the operator $(\vec\lambda \cdot \vec \nabla)\A$ on the field \eqref{mon:2} and the source \eqref{mon:3} of a point charge.  Either way, we get for the field the expression
\begin{equation}\label{dim:3}
    \vec H_{dm}(\vec r) = 
    -\vec{\nabla} \varphi_{dm}(\vec r) =
    \Bigl( 3\frac{\vec r}{r^5}(\vec\mu\cdot\vec r)
          - \frac{\vec\mu}{r^3} \Bigr)  \Upsilon(r)
  -  \frac{\vec{r}(\vec\mu\cdot\vec{r})}{r^4}\delta(r),
\end{equation}
where we have defined
\begin{equation}\label{dim:4}
                 \vec\mu = e \vec\lambda .
\end{equation}
Expression \eqref{dim:3} is remarkably similar to the corresponding expression \eqref{dip:2} for an intrinsic dipole, and it can be seen that the difference between a dipole and a dimonopole field is entirely contained in the point-like singularity at the origin, i.e., 
\begin{equation}\label{dim:9}
    \vec H_{d}(\vec r) - \vec H_{dm}(\vec r)=  
         \frac{\vec{r} \times (\vec\mu\times\vec{r})}{r^4}\delta(r) +
         \frac{\vec{r} \cdot  (\vec\mu\cdot \vec{r})}{r^4}\delta(r) =
         \frac{\vec{\mu}}{r^2}\delta(r) .
\end{equation}
As a result, when integrated over 3-space, the dimonopolar $\delta$-singular term gives the contribution \cite[p.\,141]{JACKS1975-}
\begin{equation}\label{dim:5}
       - \iiint d^3\Omega ~
         \frac{\vec{r}(\vec\mu\cdot\vec{r})}{r^4} \delta(r)
     = - \frac{4\pi}{3} \vec\mu ,
\end{equation}
which differs in sign and in magnitude from the corresponding expression \eqref{dip:3} for an intrinsic dipole.  It is this difference which enables to conclude that the dipolar fields from distant stars are produced by intrinsic magnetic dipoles, rather than by magnetic dimonopoles \cite{JACKS1977-}.

   We can now calculate the sources.  As expected, the current density is zero
\begin{equation}\label{dim:6}
    4\pi \vec j_{dm}(\vec r) = \vec\nabla \times \vec H_{dm}(\vec r) = 0 ,
\end{equation}
while the rationalized charge density is
\begin{equation}\label{dim:7}
    4\pi \rho_{dm}(\vec r) = \vec\nabla \cdot \vec H_{dm}(\vec r)
        =  3 \frac{\vec{r} \cdot \vec{\mu} }{r^4} \delta(r),
\end{equation}
i.e., a distribution that is odd in $\vec{r}$ so that the total charge is zero, as it should be for a dimonopole. We can finally calculate the \emph{first moment} of this charge density by means of the standard expression for a charge distribution \cite[p.\,137]{JACKS1975-}. This gives
\begin{equation}\label{dim:8}
            \vec d = 
     \iiint d^3\Omega ~ \vec r \rho_{dm}(\vec r)
                 = \vec\mu  =  e\vec \lambda,
\end{equation}
a result which illustrates again that despite the great similarity of their fields at a distance from the origin, the dipole and dimonopole singularities are in fact very different.

\section{Physical particles in Lanczos electrodynamics}
\label{phy:0}

The first singularity which comes to mind when considering classical electrodynamics is the electric monopole, i.e., in the non relativistic limit, the Coulomb potential
\begin{equation}\label{phy:1}
     \varphi_{\text{C}} = e\frac{1}{r}\Upsilon(r) ,
\end{equation}
and its relativistic generalization, the Li\'enard-Wiechert potential
\begin{equation}\label{phy:2}
     A_{\text{LW}} = e\frac{\mathcal{U(\tau)}}{\xi} \Upsilon(\xi) ,
\end{equation}
where the radial distance $r$ is replaced by the retarded distance $\xi$ and the 4-velocity $\mathcal{U}(\tau)$ defines the world-line of the singularity parametrized by the proper time $\tau$.  However, neither of these potentials corresponds to a physical particle since, as known from experiment, a physical electron is endowed with a magnetic dipole moment, so that the potential associated with a Dirac particle is a monopole-dipole singularity, i.e.,
\begin{equation}\label{phy:3}
     A_{\text{e}} = \varphi_{\text{monopole}} - i \vec{A}_{\text{dipole}} 
         = \Bigl( \frac{e}{r}
               - i\frac{ \vec{\mu} \times \vec{r} }{r^3} \bigr) \Upsilon(r) .
\end{equation}

   Moreover, besides from being distributions rather than ordinary functions, the potentials \eqref{phy:1} and \eqref{phy:2} present other difficulties: They lead to infinite self-energies, and to an infinite Hamiltonian function.  It is to find a cure to this second problem that in his doctoral dissertation of 1919 Lanczos proposed to shift the singularity at the origin of \eqref{phy:1} by a small imaginary translation into complex spacetime \cite[p.\,55]{LANCZ1919-}.  In this case the singularity is no more point-like but a circle in $\mathbb{C}^3$, hence the name given by Lanczos to this shifted singularity: The `circle electron.'

    Mathematically, if for simplicity we confine our discussion to the non-relativis\-tic case and to infinitesimal translations, which should be enough to get the essential features of the physical differences between the point electron and the circle electron, we can represent the shifting of the origin by the imaginary vector $i\vec\lambda$ by means of the scalar operator
\begin{equation}\label{phy:4}
     \Omega_{\text{sca}}\A = 1\Q + (i\vec\lambda\cdot\vec\nabla)\Q  .
\end{equation}
Then, applying this operator to the Coulomb potential \eqref{phy:1} we get
\begin{align}
\nonumber
     \Omega_{\text{sca}}: \varphi_{\text{C}} ~~ \rightarrow ~~
                   \varphi    &=   \Bigl(   e \frac{1}{r}
             - e(i\vec\lambda \cdot \vec r) \frac{1}{r^3}  \Bigr) \Upsilon(r)\\
\label{phy:5}
            &+  e(i\vec\lambda \cdot \vec r) \frac{1}{r^2} \delta(r),
\end{align}
i.e., a monopole-dimonopole potential on the first line, and on the second line a $\delta$-function term which can be discarded.\footnote{By convention all the \emph{potentials} considered in this paper are defined as Schwartz distributions. Hence, evaluating the $\delta$-function term in \eqref{phy:5} on any test function $T(\vec{r}) \in \mathcal{D}(\mathbb{R}^3)$ giving zero, it can be discarded.}  

   Thus, by means of $\Omega_{\text{sca}}\A$, we have added to the Coulomb potential \eqref{phy:1} an imaginary dimonopole potential of the form \eqref{dim:2}.  Since in Lanczos-Newman electrodynamics $\vec\nabla \varphi = -\vec{E} - i\vec{H}$, the effect of such an imaginary scalar potential is to produce a magnetic field which --- away from the origin --- has a dipolar form.  This led Newman to conclude that the effect of the imaginary translation \eqref{phy:4} is to induce a magnetic dipole moment \cite{NEWMA1973-,NEWMA2002-}.  However, as seen in Secs.\,\ref{sin:0}.2 and \ref{sin:0}.3, where the differences between the dipole and the dimonopole singularities were discussed, this is not the case.  In particular, the magnetic dipole field derives from a \emph{vector} potential, while the only effect of the operator \eqref{phy:4} is to make the \emph{scalar} potential \eqref{phy:1} complex instead of real.  Moreover, in the standard interpretation of Maxwell's field, imaginary contributions to the scalar potential correspond to `magnetic poles' which are normally excluded by Maxwell's inhomogeneous equations.

   What is then the physical interpretation of the monopole-dimonopole singularity \eqref{phy:5}\,?  Could there be a relation between the `dimonopolar part' of this complex scalar potential and the physics of low-energy pion-nucleon interactions which are known to be accurately described by a complex scalar field\,?  

   Let us therefore review the theory of pseudoscalar pion-nucleon interactions in its most simple form, that is in which the nucleons (i.e., protons and neutrons) are described by Dirac's equation, and the neutral pions by the scalar field $\varphi$ solution of the second order equation, \cite[p.\,9]{JACKS1958-} or \cite[p.\,436]{DEBEN1967-},
\begin{equation}\label{phy:6}
     \bigl( \nabla\CON\nabla + \ell_{\pi}^{-2} \bigr) \varphi
      = g \ell_\pi  (i\vec\nu \cdot \vec\nabla) \rho(r)  .
\end{equation}
Here $\rho(r)$ is the nuclear matter distribution source of the pion field, $\vec\nu$ the nucleon spin axis as in Dirac-Lanczos's equation \eqref{dir:1}, $\ell_\pi =  {\hbar c}/{m_\pi c^2}$ the Compton wave-length associated with the pion mass, and $g$ the strong interaction charge such that the pseudoscalar coupling constant is $g^2/\hbar c \approx 14.5$.  Assuming that the nuclear distribution is that of a point-like nucleon, i.e., $\rho(r) = \delta(r)/{4\pi r^2}$, this equation admits the static solution
\begin{equation}\label{phy:7}
\varphi(r) 
=   \frac{g}{4\pi} \ell_\pi(i\vec\nu\cdot\vec\nabla) \frac{1}{r}
    \e^{-r/\ell_\pi} \Upsilon(r)  
=  -\frac{g}{4\pi} (\ell_\pi + r) \frac{i\vec\nu\cdot\vec r}{r^3}
    \e^{-r/\ell_\pi} \Upsilon(r),
\end{equation}
which in the small pion-mass limit, $m_\pi \rightarrow 0$, or equivalently in the limit of short distances, $r \ll \ell_\pi$, reduces to
\begin{equation}\label{phy:8}
      \varphi_{(r \ll \ell_\pi)}(r) =
       -\frac{g}{4\pi} \ell_\pi \frac{i\vec\nu\cdot\vec r}{r^3}\Upsilon(r),
\end{equation}
i.e., to the dimonopole term on the right hand side of the first line of \eqref{phy:5}.  Therefore, the effect of translating the Coulomb singularity into complex space by $ {4\pi} e i\vec\lambda = {g}\ell_\pi i\vec\nu$ is equivalent to inducing a pseudoscalar interaction potential, formally similar to the dimonopole potential  \eqref{dim:2}, which can here be interpreted as corresponding to the well-known virtual pion cloud of a nucleon in low-energy pseudoscalar theory.\footnote{Of course, the action of the operator \eqref{phy:4} on the Coulomb potential $1/r$ is to lead to the pseudoscalar field \eqref{phy:5} of a \emph{massless} pion. However, the generalization to a finite-mass pion field is straightforward:  It suffice to replace $1/r$ by the Yukawa potential $\exp(-r/\ell_\pi)/r$.}

    Having shown that the \emph{scalar operator} \eqref{phy:4} is generating the strong interaction potential of a nucleon, it is natural to ask whether the magnetic interaction potential which corresponds to the vector part of the electron 4-potential \eqref{phy:3} could also be derived from the Coulomb potential by an operator of a similar type.  The answer is immediate, and it is easy to verify that \eqref{phy:3} can be derived from \eqref{phy:1} by means of the \emph{quaternion operator}
\begin{equation}\label{phy:9}
     \Omega_{\text{qua}}\A = 1\Q + (i\vec\lambda\times\vec\nabla)\Q  ,
\end{equation}
i.e.,
\begin{equation}\label{phy:10}
   \Omega_{\text{qua}}: \varphi_{\text{C}} ~~ \rightarrow ~~
    \Bigl( e \frac{1}{r}
         - e(i\vec\lambda \times \vec r)\frac{1}{r^3} \Big) \Upsilon(r)= A_{\text{e}}  ,
\end{equation}
provided $e\vec\lambda = \vec\mu$ and the $\delta$-term arising from $\Upsilon(r)$ is discarded as in \eqref{phy:5}.

   In summary, we have found that starting from the Coulomb potential --- which corresponds to a hypothetical spinless charged-particle which would have only `electric' interactions in its rest frame ---  we can derive both the magnetic-dipolar interaction potential of a physical electron, and the pseudoscalar pion-nucleon interaction potential characteristic of low-energy strong interactions, simply by applying the operators $\Omega_{\text{sca}}$ and $\Omega_{\text{qua}}$ to the world-line of the Coulomb singularity moving in biquaternionic spacetime.  Moreover, combining these results, we can immediately write down the 4-potential which in Lanczos's electrodynamics corresponds to a proton, i.e.,\footnote{This simplified expression containing only \emph{neutral} pions could easily be generalized to include charged pions.}
\begin{equation}\label{phy:11}
     A_{\text{p}} = e \frac{1}{r}
         - g (\ell_\pi + r) \frac{i\vec\nu\cdot\vec r}{r^3} \e^{-r/\ell_\pi}
         - \mu_{\text{p}} \frac{i\vec\nu\times \vec r}{r^3} ,
\end{equation}
where $\mu_{\text{p}}$ is magnetic moment of the proton.

   These results were obtained in the low-energy limit.  But there should be no reason why they would not generalize to high energies\footnote{Equation \eqref{phy:11} would then have to include other strongly interacting particles such as vector mesons, whose effect quickly become non-negligible as energy is increased \cite{SEROT1997-, MACHL2003-}.}   and to arbitrary world-lines.  The work of Newman in the general relativistic context \cite{NEWMA2004-,NEWMA1973-,NEWMA2002-},  which is essentially based on the scalar translation operator $\Omega_{\text{sca}}$, is therefore of great interest.  In particular, as stressed by Newman, it is important that the  displacement $\Omega_{\text{sca}}$ is not an ordinary Poincar\'e translation (under which all physical laws are invariant) but an imaginary translation which by complexifying a scalar potential is leading to `new' physics --- what is indeed the case since starting from a pure Maxwell field we get strong interactions.  Similarly, the operation $\Omega_{\text{qua}}$ is not an ordinary rotation or boost\footnote{Since there is some resemblance between $\Omega_{\text{qua}}$ and a rotation by an imaginary angle $i\theta$ about an axis $\vec\nu$, we recall the infinitesimal form of  such an operation acting on the coordinates, which is the scalar operator $\Omega_{\text{rot}}\A = 1\Q - \vec{r}\cdot(i\vec\lambda\times\vec\nabla)\Q$ where $\vec\lambda$ is now the dimensionless vector $\vec\lambda = \theta \vec\nu$. } (which would leave the physics invariant) but a biquaternion operation which is transforming a bare Coulomb potential into the biquaternion potential \eqref{phy:3} of a Dirac electron.  Thus, it is by these `non-orthodox' operations that we gain new insight into the nature of spacetime and elementary particles, something that was anticipated by Lanczos in 1919 already.

\section{Quantum field-theoretical and QCD perspectives}
\label{qua:0}

All considerations made so far were essentially classical, including in Sec.\,\ref{max:0} and \ref{dir:0} where the Maxwell and Dirac fields were interpreted as classical fields.  In this section we go one step further and show that when reconsidering the singularities discussed in this paper in the context of quantum field theory we can strengthen one of our main conclusions, namely that the dimonopole must be associated to strong interactions, which will therefore be valid both in the `whole particle' approach of hadrodynamics, and in the `quark gluon' approach of quantum chromodynamics (QCD).  Thus, we will show that a dimonopole is simply a `meson,' i.e., a `quark-antiquark' singularity in modern terminology.

   Indeed, as is well known, the forces mediated by any classical massless field obey the $1/r$ law, which is for instance the case of the Coulomb, Newton, and chromodynamic fields.\footnote{This is also the case of the massless pion field considered in the previous section, which however in this section is no more considered as a `fundamental' field.}  Moreover, if the properties of the gauge-groups characteristic of these fields are taken into account, it is equally well known that large modifications to the $1/r$ law arise due to the radiative corrections which come into play when these fields are treated as quantized fields \cite{GOTTF1986-}.  These modifications, when affecting the short range behavior of the strong interactions, are customarily associated with the concept of \emph{asymptotic freedom:} 
``In essence, asymptotic freedom is the statement that at very short distances the interactions between quarks is \emph{weaker} than the $1/r$ law that one would expect classically when the forces are mediated by a massless vector field.  Asymptotic freedom is a phenomenon that \emph{only} occurs in non-Abelian gauge theories (e.g., QCD).  The opposite occurs in quantum electrodynamics (QED), i.e., the interaction is stronger than $1/r$ at short distances \cite[p.\,374]{GOTTF1986-}.


   Taking these facts into account, we can revisit the concept of dimonopole singularity.  We therefore rewrite (\ref{dim:1}--\ref{dim:2}) as\footnote{This is a highly simplified picture since we ignore that the gluon field has spin and a complicated gauge group.}
\begin{equation}\label{qua:1}
 \varphi_{dm}(\vec r) =
         \frac{g_\pi}{|\vec r - \tfrac{1}{2}i\vec\lambda \,|}
       - \frac{g_\pi}{|\vec r + \tfrac{1}{2}i\vec\lambda \,|}
    \approx   ig_\pi(\vec\lambda \cdot \vec r) \frac{1}{r^3} \Upsilon(r),
\end{equation}
where we have replaced the coupling constant $e$ by $g_\pi$ because we now view the dimonopole as being a quark-antiquark singularity, and replaced $\vec{\lambda}$ by $\pm\tfrac{1}{2}i\vec{\lambda}$ in order that the potentials of the two displaced monopoles (the quark and the antiquark) are complex conjugate.

  Then, the interpretation of \eqref{qua:1} is that the dimonopole field written in that form is the field of a pseudoscalar neutral pi-meson, albeit in the limit where the radiative effects that would modify the $1/r$ law are neglected.  This form is similar to \eqref{phy:8} and confirms that the dimonopole singularity generated by the Lanczos-Newman translation must be associated with the field of a strongly interacting particle.

  Conversely, revisiting the Dirac electron singularity \eqref{phy:3}, we now better understand the truly point-like nature of both its scalar and vector parts.  In particular, the magnetic dipole is truly \emph{intrinsic} and is in no way associated to a pair of objects which like the quarks in a meson could be separated by some non-zero distance.  This applies as much to the leptons (i.e., the electron, muon, and tau) than to the quarks of all three generations, which are all true Dirac singularities.

\section{Covariant interpretation of $\Omega_{\text{sca}}$ and $\Omega_{\text{qua}}$}
\label{cov:0}

A major result of this paper is the pair of operators $\Omega_{\text{sca}}$ and $\Omega_{\text{qua}}$, defined by equations \eqref{phy:4} and \eqref{phy:9}.  They lead from the potential, field, and source of a bare monopole singularity to the distributions which characterize the physically well established pion-nucleon and photon-electron interactions.  While the scalar operator $\Omega_{\text{sca}}$ corresponds to an infinitesimal imaginary translation which has the effect of adding an imaginary scalar part to a real monopole singularity, the biquaternion operator $\Omega_{\text{qua}}$, which is adding an imaginary vector component to this monopole singularity, does not seem to have such a simple interpretation.  (Unless we associate the rotational in $\Omega_{\text{qua}}$ to a kind of vortex effect generating `spin' the same way that the divergence in $\Omega_{\text{sca}}$ relates to a translation effect generating `displacement,' so that $\Omega_{\text{qua}}$ leads to a point-like intrinsic dipole, and $\Omega_{\text{sca}}$ to a string-like dimonopole.)  Nevertheless, as will now be shown, despite this lack of a clear geometric interpretation, the two operations $\Omega_{\text{sca}}$ and $\Omega_{\text{qua}}$ are deeply connected in the context of relativity.

Indeed, to be physically meaningful $\Omega_{\text{sca}}$ and $\Omega_{\text{qua}}$ must have an invariant interpretation, which is only the case if they can be expressed in a covariant manner.  Referring to the definition of the quaternion product \eqref{int:4}, it is readily seen that the non-trivial parts of the operators $\Omega_{\text{sca}}$ and $\Omega_{\text{qua}}$ are actually the scalar and vector parts of the product
\begin{equation}\label{cov:1}
           - \bigl[\vec{\lambda}\bigr] \, \bigl[\vec{\nabla}\bigr] = 
           +  \vec{\lambda} \cdot \vec{\nabla}
           -  \vec{\lambda} \times \vec{\nabla},
\end{equation}
where the square brackets have been introduced to highlight that the product between the bracketed quantities is the quaternion product \eqref{int:4}.  The covariant formulation of \eqref{cov:1} is then obtained by replacing $-\vec\nabla$ by the quaternion-conjugate of the 4-gradient operator $\nabla$, i.e., \eqref{max:7}, and by requiring that $\vec\lambda$ transforms as a 4-vector, i.e., $\vec{\lambda}'=\mathcal{L}i\vec{\lambda}\mathcal{L}^+$ like the spin pseudo-4-vector $\Sigma$ in Dirac's theory, i.e., \eqref{dir:2}.  This yields the covariant operator
\begin{equation}\label{cov:2}
     \Omega_{\text{cov}}\A =
         i\mathcal{L}\vec{\lambda}\mathcal{L}^+\CON{\nabla}\Q,
\end{equation}
which because of the difference in signs in  \eqref{cov:1} is related to $\Omega_{\text{sca}}$ and $\Omega_{\text{qua}}$ in such a way that
\begin{equation}\label{cov:3}
     \Omega_{\text{sca}}\A - \Omega_{\text{qua}}\A = \Omega_{\text{cov}}\A .
\end{equation}
Therefore, the covariant forms of the operators $\Omega_{\text{sca}}$ and $\Omega_{\text{qua}}$ are\footnote{The operator `$\scal$' means taking the scalar part of the product of the two adjacent quaternions.}
\begin{gather}
  \label{cov:4}
    \Omega_{\text{sca}}\A  = 1\Q
 + i(\mathcal{L}\vec{\lambda}\mathcal{L}^+ \scal \CON{\nabla})\Q,\\
\label{cov:5}
    \Omega_{\text{qua}}\A  = 1\Q
 - i(\mathcal{L}\vec{\lambda}\mathcal{L}^+ \vect \CON{\nabla}) \Q,  
\end{gather}
which in the rest frame, and in the stationary case in which $\nabla$ can be replaced by $\vec{\nabla}$, yield the non-relativistic expressions \eqref{phy:4} and \eqref{phy:9}.

   In summary, the covariant treatment of $\Omega_{\text{sca}}$ and $\Omega_{\text{qua}}$ shows that these operators are related to parts of the covariant operator $\Omega_{\text{cov}}$.   Consequently, the operators $\Omega_{\text{sca}}$ and $\Omega_{\text{qua}}$ must necessarily be considered on an equal footing, which implies that the protonic and electronic interaction potentials that they generate when operating on a bare Coulomb potential are necessarily related in the same way.  That this is the case is readily seen by rewriting the corresponding operators as
\begin{gather}
\label{cov:6}
    \Omega_{\text{p}}\A \DEF \Omega_{\text{sca,p}}\A  = 1\Q
     + i\frac{g}{e}\frac{\hbar}{2m_{\text{p}} c}( \vec{\nu}  \cdot \vec{\nabla}) \Q,  \\
\label{cov:7}
    \Omega_{\text{e}}\A \DEF \Omega_{\text{qua,e}}\A  = 1\Q
        + i\frac{\hbar}{2m_{\text{e}} c}( \vec{\nu} \vect  \vec{\nabla}) \Q,  
\end{gather}
where the strong-interaction part of \eqref{cov:6} yields \eqref{phy:8} because $m_{\text{p}} = 2 \pi \, m_\pi$ to better than 10\%, and the magnetic part \eqref{cov:7} yields Dirac's magnetic dipole moment, so that the covariant superposition which generalizes \eqref{cov:3} is  ${e m_{\text{p}}}\Omega_{\text{p}}\A - {g m_{\text{e}}} \Omega_{\text{e}}\A$.

\section{Conclusion}
\label{con:0}

The possibility that Lanczos's electrodynamics and Lanczos's formulation of Dirac's theory could provide a connection between elementary particle physics and general relativity theory appears to be promising.\footnote{In this perspective it is important to stress that there are a number of approaches closely related to Lanczos's and Newman's which lead to similar conclusions.  See, in particular, \cite{KASSA2002-, BURIN1974-, BURIN2003-}.}  In particular, it is striking that in the context of a simple extension of classical electrodynamics into complex space the rather trivial Lanczos-Newman imaginary translation operation could lead to the correct low-energy pseudoscalar pion-nucleon strong interaction, and that by a fully covariant treatment the intrinsic magnetic dipolar interaction of a Dirac particle would necessarily arise on an equal footing, i.e., from (\ref{cov:6}--\ref{cov:7}),  
\begin{align}
\label{con:1}
 \Omega_{\text{p}}: +e \frac{1}{r} ~~ \rightarrow ~~
    &A_{\text{p}} = +\Bigl( e \frac{1}{r}  - i\frac{g\hbar}{2m_{\text{p}} c}
        \frac{\vec\nu  \cdot \vec r}{r^3} \Big),\\
\label{con:2}
 \Omega_{\text{e}}: -e \frac{1}{r} ~~ \rightarrow ~~
    &A_{\text{e}} = -\Bigl( e \frac{1}{r} - i\frac{e\hbar}{2m_{\text{e}} c}
        \frac{\vec\nu \times \vec r}{r^3} \Big).
\end{align}
This could mean that the well-known formal similarities between magnetodynamics and low-energy strong interactions, see \cite[p.\,3--10]{JACKS1958-} and \cite[p.\,434--439]{DEBEN1967-}, are not just accidental but of a fundamental nature, which may be related to elementary geometric operations on the world lines of monopolar singularities moving in complexified spacetime.

\section{Acknowledgments}
\label{ack:0}

This paper is dedicated to Prof.\ J.D. Jackson whose enlightening work on the nature of intrinsic magnetic dipole moments has greatly influenced the results presented here.  I am very much indebted to Prof.\ E.T. Newman for an extensive correspondence, as well as to Dr.\ J.-P. Hurni for considerable assistance in several aspects of this paper.


\section{References}
\label{biblio:0}


\begin{enumerate} 


\bibitem{NEWMA2004-} E.T. Newman, \emph{Maxwell fields and shear-free null geodesic congruences}, Class. Quant. Grav. {\bf 21} (2004) 3197--3222. e-print arXiv:gr-qc/0402056.

\bibitem{NEWMA1973-} E.T. Newman, \emph{Maxwell's equations and complex Minkowski space}, J. Math. Phys. {\bf 14} (1973) 102--103. 

\bibitem{NEWMA2002-}  E.T. Newman, \emph{Classical, geometric origin of magnetic moments, spin-angular momentum, and the Dirac gyromagnetic ratio}, Phys. Rev. {\bf D65} (2002) 104005-1/8. e-print arXiv:gr-qc/0201055.

\bibitem{SEROT1997-} B.D. Serot and J.D. Walecka, \emph{Recent progress in quantum hadrodynamics}, Int. J. Mod. Phys. {\bf E6} (1997) 515--631. e-print arXiv:nucl-th/9701058.

\bibitem{MACHL2003-} R. Machleidt and D.R. Entem, \emph{The nuclear force problem: Are we seeing the end of the tunnel~?} Lead talk presented at the 17th International Conference on Few-Body Problems in Physics (Duke University, June 2003) 5pp.  e-print arXiv:nucl-th/0309026.

\bibitem{GSPON2001-} A. Gsponer, \emph{Comment on Formulating and Generalizing Dirac's, Proca's, and Maxwell's Equations with Biquaternions or Clifford Numbers}, Found. Phys. Lett. {\bf 14} (2001) 77--85. e-print  arXiv:math-ph/0201049.

\bibitem{LANCZ1919-} Kornel Lanczos, Die Funktionentheoretischen Beziehungen der Max\-well\-schen  Aethergleichungen  --- Ein Beitrag zur  Relativit\"ats- und   Elektronentheorie (Verlagsbuchhandlung  Josef   N\'emeth, Budapest, 1919) 80\,pp.

Lanczos's manuscript dissertation is reprinted in the Appendix of W.R. Davis \emph{et al.}, eds., Cornelius Lanczos Collected Published Papers With Commentaries, {\bf 6} (North Carolina State University, Raleigh, 1998) pages A-1 to A-82. Web site  http://www.physics.ncsu.edu/lanczos.  

Lanczos's  dissertation is also available in typeseted form as:

\bibitem{HURNI2004-} Cornelius Lanczos, \emph{The relations of the homogeneous Maxwell's equations to the theory of functions --- A contribution to the theory of relativity and electrons} (1919, Typeseted by Jean-Pierre Hurni with a preface by Andre Gsponer, 2004); e-print arXiv:physics/0408079.

\bibitem{GSPON1998A} A. Gsponer and J.-P. Hurni, \emph{Lanczos's functional theory of electrodynamics --- A commentary on Lanczos's PhD dissertation,} {\bf in} W.R. Davis \emph{et al.}, eds., \emph{op. cit.}, {\bf 1} (1998) pages 2-15 to 2-23. e-print arXiv:math-ph/0402012.

\bibitem{GSPON2004E}  A. Gsponer and J.-P. Hurni, \emph{Cornelius Lanczos's derivation of the usual action integral of classical electrodynamics}, Found. Phys. {\bf 35} (2005) 865--880.  e-print arXiv:math-ph/0408027

\bibitem{LANCZ1929A} C. Lanczos, \emph{The tensor analytical relationships of Dirac's equation}, Zeits. f. Phys. {\bf 57} (1929) 447--473. Reprinted and translated {\bf in} W.R. Davis \emph{et al.}, eds., \emph{op. cit.}, {\bf 3} (1998) pages 2-1132 to 2-1185.  e-print  arXiv:physics/0508012.

\bibitem{LANCZ1929B} C. Lanczos, \emph{The covariant formulation of Dirac's equation}, Zeits. f. Phys. {\bf 57} (1929) 474--483. Reprinted and translated {\bf in} W.R. Davis \emph{et al.}, eds., \emph{op. cit.}, {\bf 3} (1998) pages 2-1186 to 2-1205.  e-print arXiv:physics/0508002.

\bibitem{LANCZ1929C} C. Lanczos, \emph{The conservation laws in the field theoretical representation of Dirac's theory}, Zeits. f. Phys. {\bf 57} (1929) 484--493. Reprinted and translated {\bf in} W.R. Davis \emph{et al.}, eds., \emph{op. cit.}, {\bf 3} (1998) pages 2-1206 to 2-1225.  e-print  arXiv:physics/0508009.

\bibitem{GSPON1998B} A. Gsponer and J.-P. Hurni, \emph{Lanczos-Einstein-Petiau: From Dirac's equation to non-linear wave mechanics,} {\bf in} W.R. Davis \emph{et al.}, eds., \emph{op. cit.}, {\bf 3} (1998) pages 2-1248 to 2-1277.  e-print  arXiv:physics/0508036.

\bibitem{GSPON1993-} A.  Gsponer and J.-P.  Hurni,  \emph{The Physical Heritage of W.R. Hamilton.}  Lecture at the conference `The Mathematical Heritage of  Sir William Rowan Hamilton,' 17-20 August, 1993, Dublin, Ireland.\\  e-print 
arXiv:math-ph/0201058.

\bibitem{GSPON1994-} A. Gsponer and J.-P. Hurni, \emph{Lanczos's equation to replace Dirac's equation~?} {\bf in} J.D. Brown et al., eds, Proc. Int. Cornelius Lanczos Conf., Raleigh, NC, USA (SIAM Publ., 1994) 509--512.  There is an error and a number misprints in this paper.  Please refer to the corrected version, e-print arXiv:hep-ph/0112317.

\bibitem{GSPON2002-} A. Gsponer, \emph{On the `equivalence' of the Maxwell and Dirac equations}, Int. J. Theor. Phys. {\bf 41} (2002) 689--694.  e-print arXiv:math-ph/0201053.

\bibitem{GURSE1958-} F. G\"ursey, \emph{Relation of charge independence and baryon conservation to Pauli's transformation}, Nuovo Cim. {\bf 7} (1958) 411--415.

\bibitem{GURSE1956B} F. G\"ursey, \emph{Correspondence between quaternions and four-spinors}, Rev. Fac. Sci. Istanbul {\bf A 21} (1956) 33--54.

\bibitem{TANGH1962-} F.R. Tangherlini, \emph{General relativistic approach to the Poincar\'e compensating stresses for the classical point electron}, Nuovo Cim. {\bf 26} (1962) 497--524.

\bibitem{GSPON2004D} A. Gsponer, \emph{Distributions in spherical coordinates with applications to classical electrodynamics}, Eur. J. Phys. {\bf 28} (2007) 267--275; Corrigendum Eur. J. Phys. {\bf 28} (2007) 1241. e-print arXiv:physics/0405133.

\bibitem{GSPON2006B} A. Gsponer, \emph{A concise introduction to Colombeau generalized functions and their applications to classical electrodynamics} (2006) 19\,pp. To be published in Eur. J. Phys. e-print arXiv:math-ph/0611069.

\bibitem{GSPON2008B} A. Gsponer, \emph{The classical point-electron in Colombeau's theory of generalized functions}, J. Math. Phys. {\bf 49} (2008) 102901 \emph{(22 pages)}. e-print arXiv:0806.4682.

\bibitem{JACKS1975-} J.D. Jackson, Classical electrodynamics (J. Wiley \& Sons, New York, second edition, 1975) 848\,pp.

\bibitem{JACKS1977-} J.D. Jackson, \emph{On the nature of intrinsic magnetic dipole moments}, CERN report 77-17 (CERN, Geneva, 1 Sept. 1977) 18 pp. Reprinted {\bf in} V. Stefan and V.F. Weisskopf, eds., Physics and Society: Essays in Honor of Victor Frederick Weisskopf (AIP Press, New York, Springer, Berlin, 1998)  129--152.

\bibitem{JACKS1958-} J.D. Jackson, The Physics of Elementary Particles (Princeton University Press, Princeton, 1958) 135\,pp.

\bibitem{DEBEN1967-} S. DeBenedetti, Nuclear Interactions (J. Wiley \& Sons, New York, 1967) 636\,pp.

\bibitem{GOTTF1986-} K. Gottfried and V.F. Weisskopf, Concepts in Particle Physics, Vol.\,II (Oxford University Press, 1986) 608\,pp. 

\bibitem{STEPH1957-} G. Stephenson, \emph{A classical calculation of the nucleon-meson coupling constant}, Nuovo Cimento {\bf 5} (1957) 1009--1010.

\bibitem{KASSA2002-} V.V. Kassandrov, \emph{General solution of the complex 4-eikonal equation and the `algebrodynamical' field theory}, Gravitation \& Cosmology {\bf 8}, Suppl.2 (2002) 57--62. e-print arXiv:math-ph/0311006.

\bibitem{BURIN1974-} A.Ya. Burinskii, \emph{Microgeons with spin}, Sov. Phys. JETP {\bf 39} (1974) 193--195.

\bibitem{BURIN2003-} A.Ya.  Burinskii, \emph{Complex Kerr geometry and nonstationary Kerr solutions}, Phys. Rev. {\bf D 67} (2003) 124024. e-print arXiv:gr-qc/0212048.


\end{enumerate}
\end{document}